\documentclass[%
preprint%
 ,secnumarabic%
,amssymb, amsmath,nobibnotes,aps]{revtex4}

\usepackage{epsfig}%
\usepackage{graphicx}%
\expandafter\ifx\csname package@font\endcsname\relax\else
 \expandafter\expandafter
 \expandafter\usepackage
 \expandafter\expandafter
 \expandafter{\csname package@font\endcsname}%
\fi

\begin{document}

\title{Generation of scale invariant density perturbations in a conformally invariant Inert Higgs doublet model}
\author{Moumita Das}
\email{moumita@prl.res.in}
\affiliation{Physical Research Laboratory, Ahmedabad 380009,
India}
\author{Subhendra Mohanty}
\email{mohanty@prl.res.in}
\affiliation{Physical Research Laboratory, Ahmedabad 380009,
India}
\def\be{\begin{equation}}
\def\ee{\end{equation}}
\def\al{\alpha}
\def\bea{\begin{eqnarray}}
\def\eea{\end{eqnarray}}

\begin{abstract}
If a Higgs field is conformally coupled to gravity, then it can give rise to the scale invariant density
perturbations. We make use of this result in a realistic inert Higgs doublet model, where we have a pair
of Higgs doublets  conformally coupled to the gravity in the early universe. The perturbation of the inert
Higgs is shown to be the scale invariant. This gives rise to the density perturbation observed through CMB
by its couplings to the standard model Higgs and the subsequent decay. 
Loop corrections of this conformally coupled system gives rise to electroweak symmetry breaking.
We constrain the couplings of the scalar potential by comparing with the amplitude and spectrum of CMB
anisotropy measured by WMAP and this model leads to a prediction for the masses of the lightest Higgs
and the other scalars.
\end{abstract}
\maketitle

\section{Introduction}

It is well known that to generate the density perturbation of the CMB of the magnitude observed by COBE and WMAP,
we need an inflationary period generated by the flat potential of a scalar field with
coupling $\lambda \sim 10^{-10}$ in a $\lambda \phi^4$ theory. For standard model Higgs, $\lambda$ is
approximately $\sim 1 $ and the Higgs can not be used as inflaton. A way out was proposed  by Bezrukov
and Shaposhnikov \cite{Bezrukov:2007ep} who coupled the standard model Higgs with the Ricci scalar
with a large coupling constant $\xi\sim 10^4$. This large coupling leads to problem
with unitarity \cite{Hertzberg:2010dc,Burgess:2010zq,David} of graviton-scalar scattering. Some attempts to solve the
unitarity problem associated with the large Higgs curvature coupling are in
\cite{Germani:2010gm,Lerner:2009na,Giudice}

In this paper we follow a different approach for the generation of scale invariant density perturbations. It was shown
by Rubakov and collaborators \cite{Rubakov:2009np,Libanov:2010nk,Libanov:2011hh} that a conformally coupled field
rolling down a quartic potential can generate scale invariant density perturbation. These perturbations can become
superhorizon in an inflationary era or in a ekpyrotic scenario \cite{Khoury}. We work with the inert Higgs doublet
(IDM) model \cite{Barbieri:2006dq,Hambye:2007vf} with conformal couplings to the Ricci scalar. The mass terms which
give rise to electroweak symmetry breaking are generated by the  Coleman-Weinberg method ~\cite{Coleman:1973jx}.

The requirement of scale invariance at high energy scale and electro-weak symmetry breaking at low energies fixes
the coupling constants of the theory. Specifically we find that the quartic coupling of the inert doublet,
 predicts a spectral index of the power spectrum of the perturbations to be consistent with observations. The
amplitude of the power spectrum $P_\zeta$ can be tuned to be consistent with the observations by choosing a
suitable curvaton mechanism. This model specifies the mass of the Higgs boson to be $m_h=291$ GeV and the mass
of the dark matter $m_{A_0}=550$ GeV which can be tested respectively at the LHC and in cosmic ray observations.
The main aim of this paper has been to show that a Higgs potential with not too small couplings can be a viable 
source of the observed scale invariant density perturbations. The scale invariant density perturbations become 
superhorizon during a phase of inflation at the electroweak scale. However other cosmological scenarios like a 
bounce models \cite{Khoury} of making the density perturbations superhorizon may be equally viable with our model.

In Section-2 we describes the basics of the Inert Doublet Model (IDM). The one-loop correction to the potential and
the calculation for running of coupling constant are briefly discussed in Section-3. We study the generation of
the scale invariant density perturbation from the inert higgs doublet in Section-4. In Section-5 we list the scalar
mass spectrum predicted by this model and identify the dark matter candidate.


\section{Inert Doublet Model}
Inert Doublet Model (IDM) is a economical extension of Standard Model which solves the problem of naturalness
\cite{Barbieri:2006dq} and it can also explain the electroweak symmetry breaking \cite{Hambye:2007vf}.
 The
lagrangian of this model respect the $Z_2$ symmetry, under which all Standard model particles including
the SM Higgs $H_1$ are even and an extra scalar doublet $H_2$ is odd.
Due to $Z_2$ symmetry, the cubic term and yukawa term for $H_2$ doublet are forbidden. This makes the inert
doublet stable and its neutral component can be a candidate for dark matter. The two Higgs doublets $H_1$ and
$H_2$ can be written in terms of their component fields as,\\
\begin{eqnarray}
 H_1=\left(
\begin{array}{l}
h^{+} \\
\frac{h+ i G_0 }{\sqrt{2}}
\end{array}
\right)
\qquad\quad
H_2= \left(
\begin{array}{l}
H^{+} \\
\frac{H_0+ i A_0 }{\sqrt{2}}
\end{array}
\right)\nonumber
\end{eqnarray}\\

The most general renormalisable potential  will be,
\begin{eqnarray}
V_{tree}=V_c+\mu_1|H_1|^{2}+\mu_2 |H_2|^{2}+\lambda_1|H_1|^{4}+\lambda_2 |H_2|^{4}+
\lambda_3 |H_1|^{2} |H_2|^{2}+\nonumber \\
\lambda_4 |H_1^{\dagger}H_2|^{2}+\frac{\lambda_5}{2}\left[\left(H_1^{\dagger}H_2\right)^{2}+h.c.\right]
\label{Vtree}
\end{eqnarray}
We consider the conformal case where $\mu_1\,=\,\mu_2=0$. $V_c$ is the constant potential, which acts as cosmological
constant and can be formed from the vev of different Higgs fields. We have chosen $V_c=3.66\times10^8\,\rm{GeV}^4$ such that
the minimum of the total potential becomes zero at present era. In the early universe the cosmological constant gives
rise to an exponential expansion during which the scale invariant perturbations of the phase of the neutral component
of $H_2$ become super-horizon. To achieve this we need that the potential is such that in the early universe,
$V \sim -|\lambda_2| |H_2|^4 $ and the neural component of $H_2$ rolls down this quartic potential while the minimum
of $H_1$ is at $\langle H_1 \rangle=0$. In the present era the potential should be such that the minima occurs at
$\langle H_2 \rangle=0$ and $\langle H_1\rangle=v=246 GeV$ which gives rise to the electro-weak symmetry breaking.
We show in the next section how this is achieved by radiative corrections starting from a scale invariant tree level potential.

\section{Coleman-Weinberg loop correction}

We derive the one-loop correction to the potential (\ref{Vtree}) following Coleman-Weinberg formalism
~\cite{Coleman:1973jx}. The generic one-loop correction to the potential can be written as ~\cite{Mooij:2011fi},\\
\begin{eqnarray}
\Delta V^{1}=\frac{1}{2}\sum_i (-1)^{2J_i}\,\left(2J_i+1\right)\,\int\frac{d^3 k}{(2\pi)^3}\,\sqrt{k^2+m_i^2}
\label{divergent}
\end{eqnarray}
where $J_i$ is the spin of the fields and $m_i$ are the tree level masses, function of the Higgs field. The double
derivative of the tree level potential (\ref{Vtree}) with respect to the fields give the the tree level masses, which are,
\begin{eqnarray}
m_h^2&=&\lambda_1 (G_0^2+3h^2+2h^{+}h^{-})+\lambda_3 H^{+}H^{-}+\frac{\lambda_L}{2} H_0^2+\frac{\lambda_S}{2} A_0^2\nonumber\\
m_{G_0}^2&=&\lambda_1 (h^2+3G_0^2+2h^{+}h^{-})+\lambda_3 H^{+}H^{-}+\frac{\lambda_L}{2} A_0^2+\frac{\lambda_S}{2} H_0^2\nonumber\\
m^2_{h^{\pm}}&=&2\lambda_1 (G_0^2+h^2+6h^{+}h^{-})+\lambda_3 ( H_0^2+A_0^2)+2\lambda_L H^{+}H^{-}\nonumber\\
m_{H_0}^2&=&\lambda_2 (A_0^2+3H_0^2+2H^{+}H^{-})+\lambda_3 h^{+}h^{-}+\frac{\lambda_L}{2} h^2+\frac{\lambda_S}{2} G_0^2\nonumber\\
m_{A_0}^2&=&\lambda_2 (H_0^2+3A_0^2+2H^{+}H^{-})+\lambda_3 h^{+}h^{-}+\frac{\lambda_L}{2} G_0^2+\frac{\lambda_S}{2} h^2\nonumber\\
m^2_{H^{\pm}}&=&2\lambda_2 (A_0^2+H_0^2+6H^{+}H^{-})+\lambda_3 ( h^2+G_0^2)+2\lambda_L h^{+}h^{-}
\label{t_l_mass}
\end{eqnarray}
where $\lambda_{L,S}\equiv\lambda_3 +\lambda_4 \pm\lambda_5$.
We regularize the divergent terms in
Eq.~(\ref{divergent}) using the cut-off scale $\Lambda$ and obtain
\begin{eqnarray}
\Delta V^{1}=\sum_i\,\left(\frac{m_i^2 \Lambda^2}{32 \pi^2}+\frac{m_i^4}{64 \pi^2} \left(\ln\frac{m_i^4}{\Lambda^2}-\frac{1}{2}\right)\right)
\label{divergence}
\end{eqnarray}
The divergence in Eq.~(\ref{divergence}) can be removed by adding the counter terms in the potential of the form,
\begin{eqnarray}
V_{ct}(\phi)=\delta \mu_{\phi}^2\phi^2+\delta\lambda_{\phi}\phi^4
\end{eqnarray}
where $\phi$ denote the scalar fields, considered in the model.

We impose the regularization condition on the effective potential, such that at early era (with high $\mu$ value), the
potential is scale invariant form (\ref{Vtree}) by choosing the counter terms as follows,\\
\begin{eqnarray}
\delta \mu_{\phi}^2\phi^2 &=& \left(6\lambda_1+2\lambda_3+\lambda_4+1/2\right)\,h^2+\left(6\lambda_2+2\lambda_3+\lambda_4\right)\,A^2\nonumber\\
&+&\left(6\lambda_1+2\lambda_3+\lambda_4\right)\,G^2 +\left(6\lambda_2+2\lambda_3+\lambda_4\right)\,H^2\nonumber\\
&+&2\left(8\lambda_1+2\lambda_3+\lambda_4+\lambda_5\right)\,h^{+}h^{-}+2\left(8\lambda_2+2\lambda_3+\lambda_4+\lambda_5\right)\,H^{+}H^{-}
\end{eqnarray}
\begin{eqnarray}
\delta\lambda_{\phi}\phi^4 &=& h^4\left(9\lambda_1^2\,f(m_h^2)+\lambda_1^2\,f(m_{G_0}^2)+4\lambda_1^2\,f(m_{h^{\pm}}^2)+\frac{\lambda_L^2}{4}\,f(m_{H_0}^2)+\frac{\lambda_S^2}{4}\,f(m_{A_0}^2)+\lambda_3^2\,f(m_{H^{\pm}}^2)\right)\nonumber\\
&+& H_0^4\left(\frac{\lambda_L^2}{4}\,f(m_h^2)+\frac{\lambda_S^2}{4}\,f(m_{G_0}^2)+\lambda_3^2\,f(m_{h^{\pm}}^2)+9\lambda_2^2\,f(m_{H_0}^2)+\lambda_2^2\,f(m_{A_0}^2)+4\lambda_2^2\,f(m_{H^{\pm}}^2)\right)\nonumber\\
&+& G_0^4\left(\lambda_1^2\,f(m_h^2)+9\lambda_1^2\,f(m_{G_0}^2)+4\lambda_1^2\,f(m_{h^{\pm}}^2)+\frac{\lambda_3^2}{4}\,f(m_{H_0}^2)+\frac{\lambda_L^2}{4}\,f(m_{A_0}^2)+\lambda_3^2\,f(m_{H^{\pm}}^2)\right)\nonumber\\
&+& A_0\left(\frac{\lambda_S^2}{4}\,f(m_h^2)+\frac{\lambda_L^2}{4}\,f(m_{G_0}^2)+\lambda_3^2\,f(m_{h^{\pm}}^2)+\lambda_2^2\,f(m_{H_0}^2)+9\lambda_2^2\,f(m_{A_0}^2)+4\lambda_2^2\,f(m_{H^{\pm}}^2)\right)\nonumber\\
&+& (h^{+}h^{-})^2\left(4\lambda_1^2\,f(m_h^2)+4\lambda_1^2\,f(m_{G_0}^2)+144\lambda_1^2\,f(m_{h^{\pm}}^2)\right.\nonumber\\
&+&\left.\lambda_3^2\,f(m_{H_0}^2)+\lambda_3^2\,f(m_{A_0}^2)+4\lambda_L^2\,f(m_{H^{\pm}}^2)\right)\nonumber\\
&+& (H^{+}H^{-})^2\left(\lambda_3^2\,f(m_h^2)+\lambda_3^2\,f(m_{G_0}^2)+4\lambda_L^2\,f(m_{h^{\pm}}^2)\right.\nonumber\\
&+&\left.4\lambda_L^2\,f(m_{H_0}^2)+4\lambda_2^2\,f(m_{A_0}^2)+144\lambda_L^2\,f(m_{H^{\pm}}^2)\right)
\end{eqnarray}\\
where $f(m_i^2)= \log\left(\frac{\Lambda^2}{\mu^2}+\frac{\mu^2}{m_i^2}\right)$.
With these counter terms the form of the effective potential turns out to be,\\
\begin{eqnarray}
V_{eff.}(H,h,\mu)= V_{tree}+\frac{1}{64\pi^2}\sum_i n_i m_i^4 \ln(\frac{m_i^2}{\mu^2}+1)
\label{eff_pot}
\end{eqnarray}\\
where $n_i$ is degrees of freedom and $m_i$ are tree level masses, shown in Eq.~(\ref{t_l_mass}).

To get the correct electro-weak symmetry breaking in the present era and the scale invariant density perturbation in the
early era, we have chosen a set of $\lambda$ values in present epoch as shown in Table-(\ref{table1}).

The values of $\lambda_i$, where $\{i=3\,\rm{to}\, 5\}$ are chosen such that we can get electro-weak symmetry breaking in the present
era. 
\begin{center}
\begin{table}[h]
\begin{tabular}{|c|c|c|c|c|}\hline
$\lambda_1$ & $\lambda_2$  & $\lambda_3$ & $\lambda_4$ & $\lambda_5$   \\
\hline
$\,\,$-1.28 $\,\,$& $\,\,$-21$\,\,$& $\,\,$5.9 $\,\,$ & $\,\,$-2.8$\,\,$ & $\,\,$-2.8$\,\,$  \\
\hline
\end{tabular}
\caption{The scalar couplings in the present era with $\mu=172.5$ GeV}
\label{table1}
\end{table}
\end{center}
\begin{figure}[ht]
\begin{center}
\includegraphics[height=6.5cm,width=9.5cm]{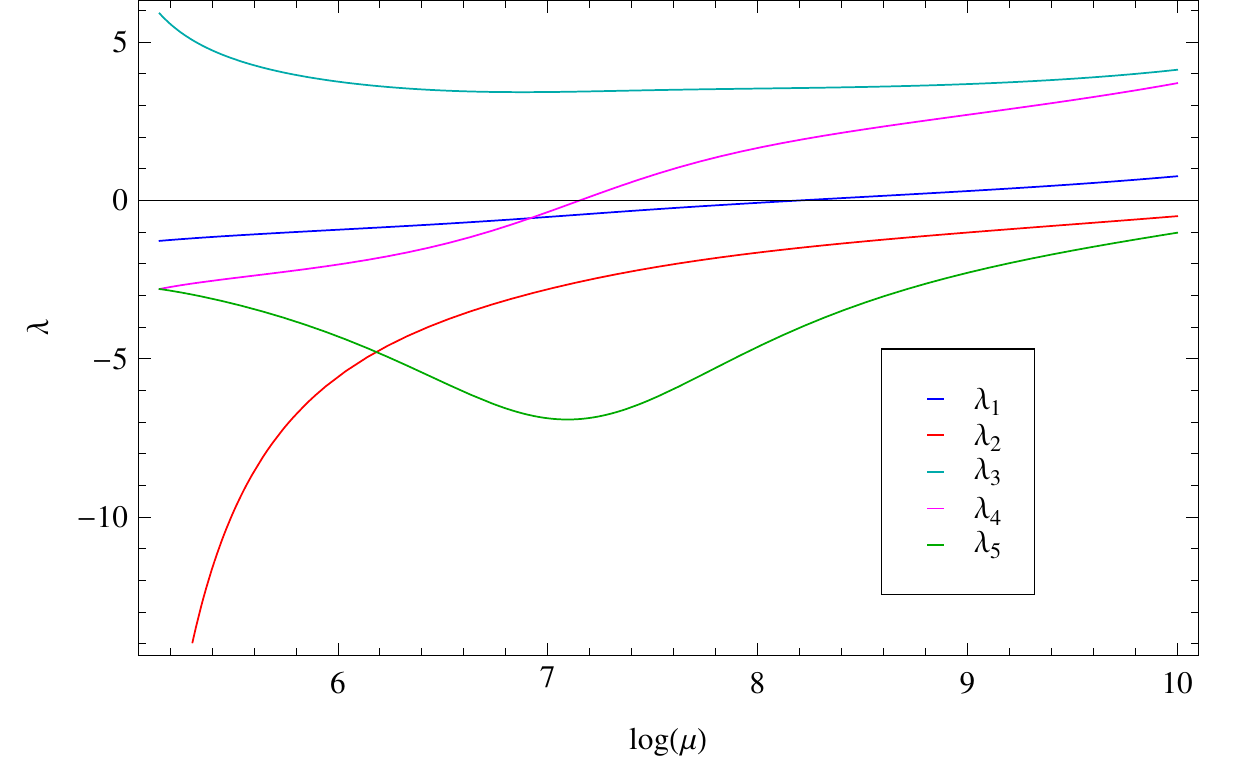}
\caption{Running of $\lambda$ from present to early era }
\label{lambda_1}
\end{center}
\end{figure}

Now we have study the running of couplings $\lambda_i$, where $\{i=1\,\rm{to}\, 5\}$ using the one-loop renormalization group equation
for the inert doublet model \cite{Barbieri:2006dq}.

From Fig~(\ref{lambda_1}), we can find the $\lambda$ values in the early era $\mu\simeq 10^4$ are as follows,
\begin{center}
\begin{table}[h]
\begin{tabular}{|c|c|c|c|c|}\hline
$\lambda_1$ & $\lambda_2$  & $\lambda_3$ & $\lambda_4$ & $\lambda_5$   \\
\hline
$\,\,$0.76 $\,\,$& $\,\,$-0.5$\,\,$& $\,\,$4.12 $\,\,$ & $\,\,$3.70$\,\,$ & $\,\,$-1.02$\,\,$  \\
\hline
\end{tabular}
\caption{The scalar couplings in the early era with $\mu=10^4$ GeV}
\label{table2}
\end{table}
\end{center}
Only $\lambda_2$ at early universe is relevant for calculating the scale invariant density perturbation.

\begin{figure} [h]
\begin{center}
\includegraphics[height=0.35\textwidth]{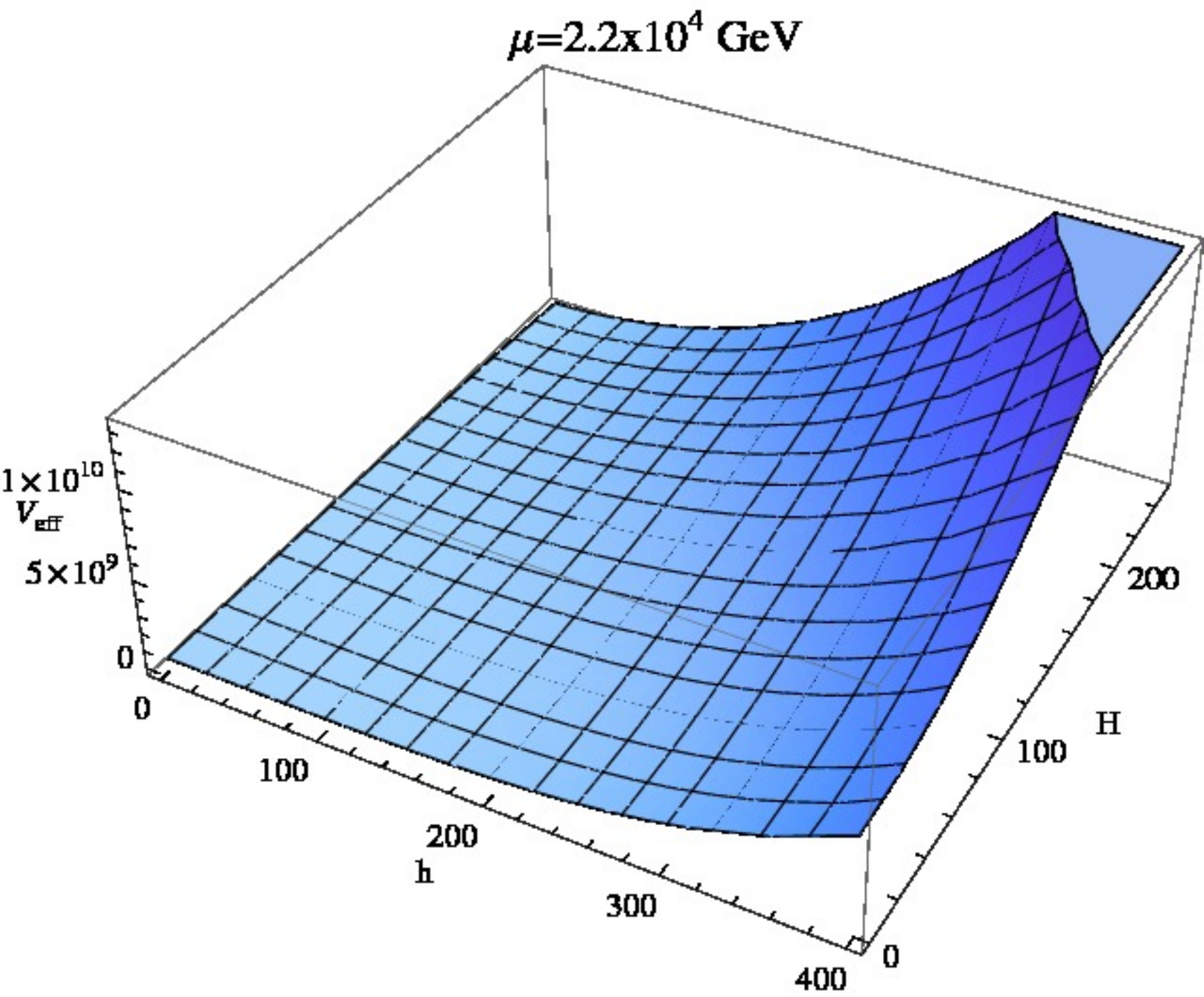}\\
\includegraphics[height=0.35\textwidth]{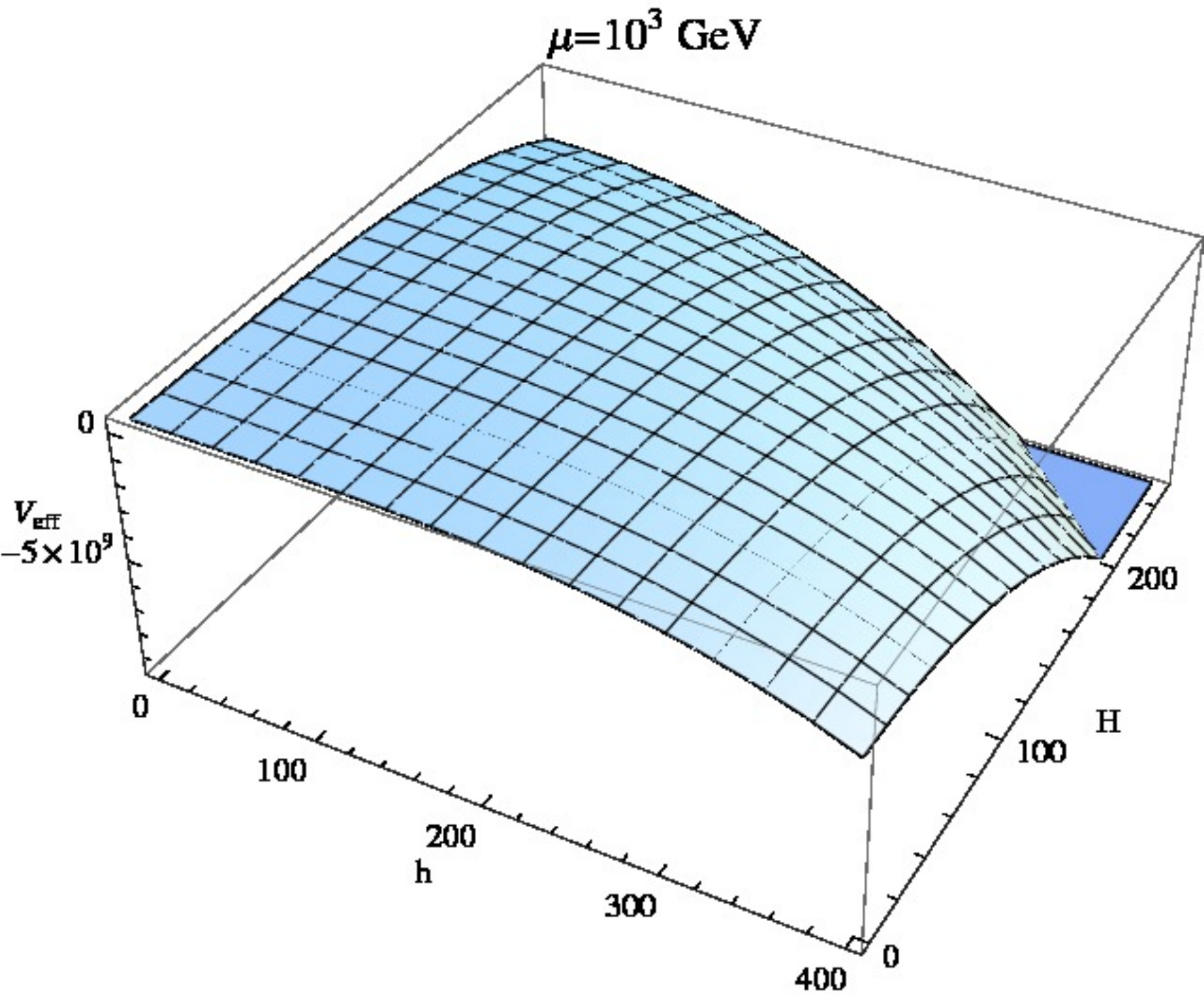}\\
\includegraphics[height=0.35\textwidth]{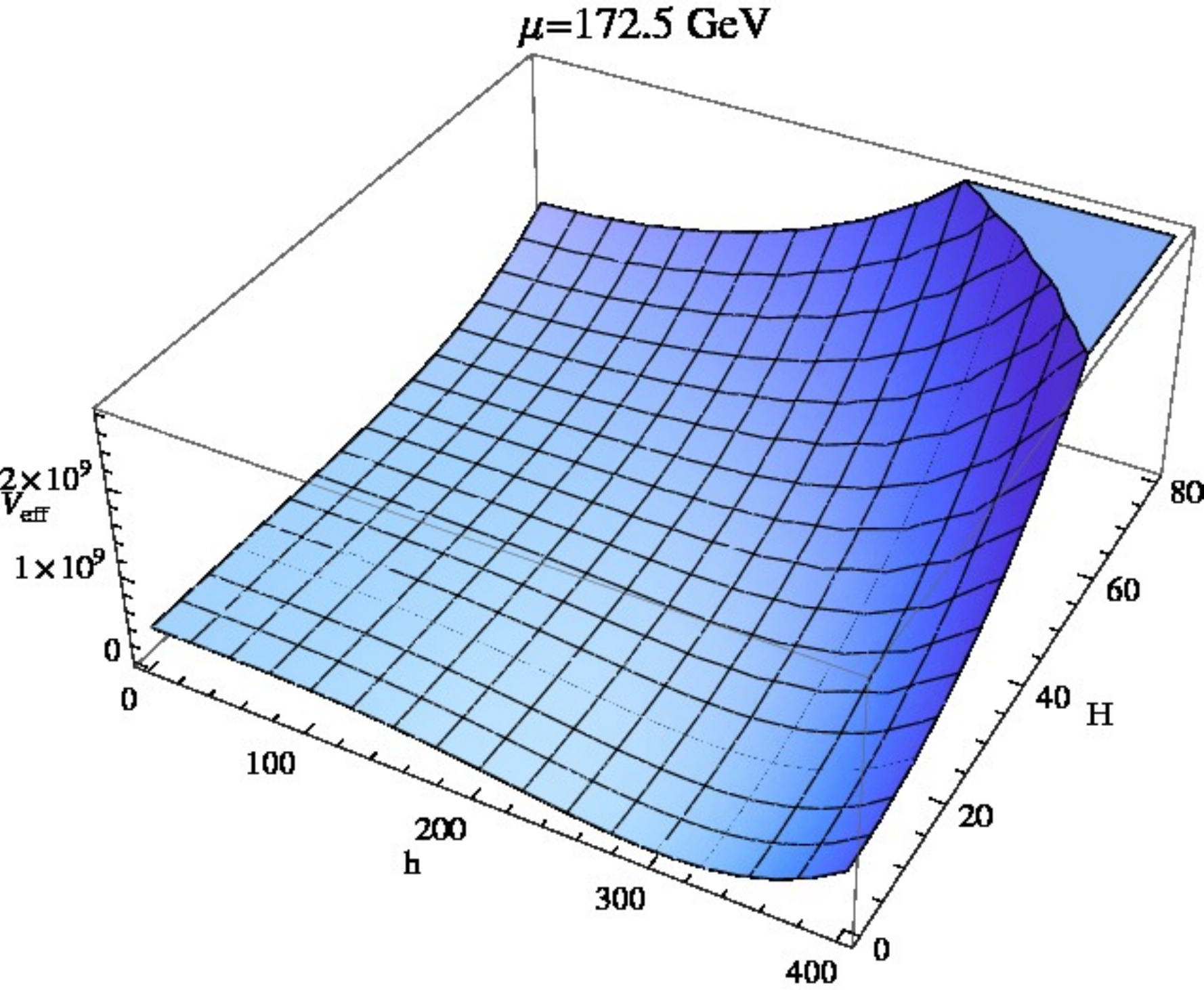}\\
\caption{Variation of the potential at different era}
\label{plot-3D}
\end{center}
\end{figure}

The change in shape of the  effective potential $V_{eff}(H,h,\mu)$ in Eq.~(\ref{eff_pot}) from the early universe
where we take $\mu=2.2\times10^{4}$ to the present epoch where $\mu=172.5$ GeV is shown in Fig~(\ref{plot-3D}).
We see that in the early universe for a given value of $H$ the minima of $V_{eff}(h)$ (shown in Fig 3(a)) is at $h=0$.
We assume that in the early universe $h=0$ and we see that $V_{eff}(H)$ is of the form as shown in Fig 3(b).

The one loop correction of the potential has significance contribution in present era. When we take $\mu=172$
GeV then the potential (\ref{eff_pot}) is of the form shown in Fig 4(a) and Fig 4(b). In this era, the $V_{eff}(H)$
has a minimum at $H=0$ as shown in Fig 4(b). With $H=0$ , the potential as a function of the field $h$ has a
minimum at $h=v \sim 246 GeV$ signifying the electroweak symmetry breaking.
We note that the potential we calculate are at zero temperature which accurately describes the universe during inflation 
(when any prior temperature goes down exponentially in time) or in the present universe where the background temperature 
negligible compared to the electroweak scale. There is a radiation era after  re-heating at the end of inflation. 
The effective potential at high temperature has been computed for the inert Higgs doublet model in 
\cite{Chowdhury:2011ga}, where the thermal evolution of the effective potential has been shown. In this paper we deal with 
the $T=0$ case which is relevant during inflation and in the present universe.

\begin{figure} [h]
\begin{center}
\includegraphics[height=0.30\textwidth]{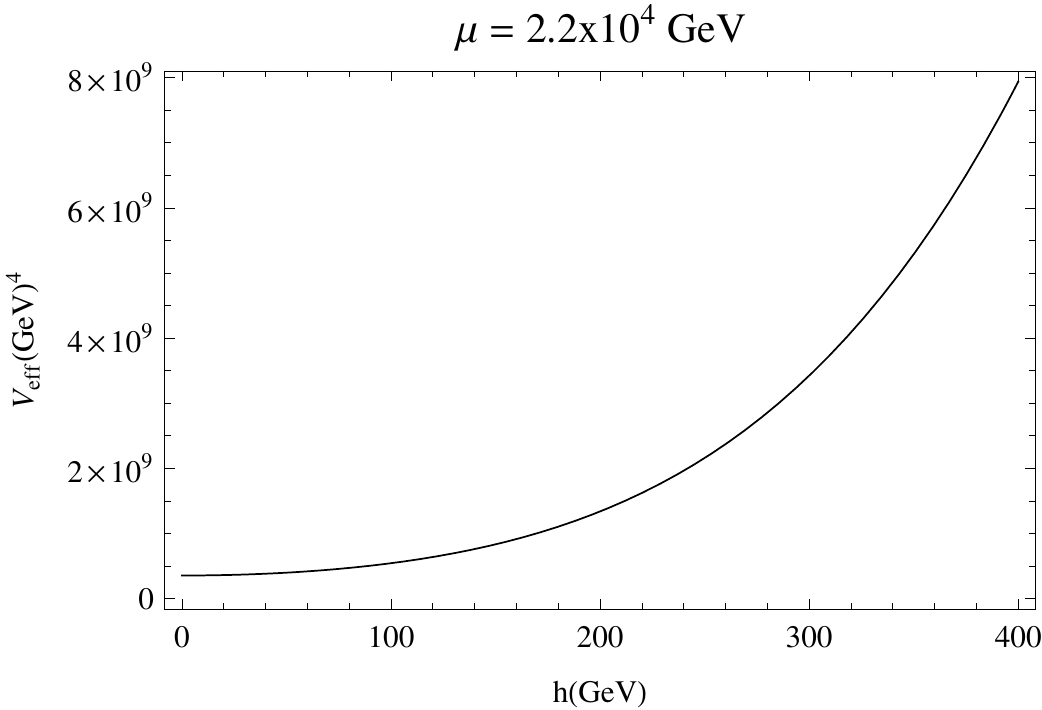}
\hskip20pt
\includegraphics[height=0.30\textwidth]{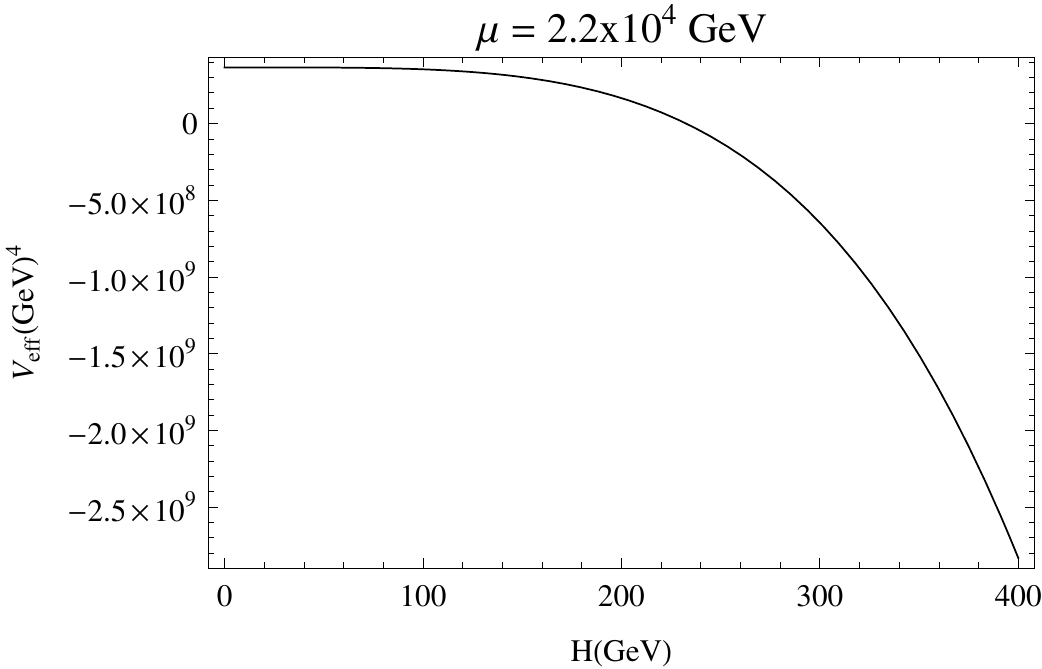}
\vspace*{0.2cm}
\hskip5pt  (a)
\hskip240pt  (b)
\caption{Effective potential  in the early universe}
\label{early_plot}
\end{center}
\end{figure}

\begin{figure} [h]
\begin{center}
\includegraphics[height=0.30\textwidth]{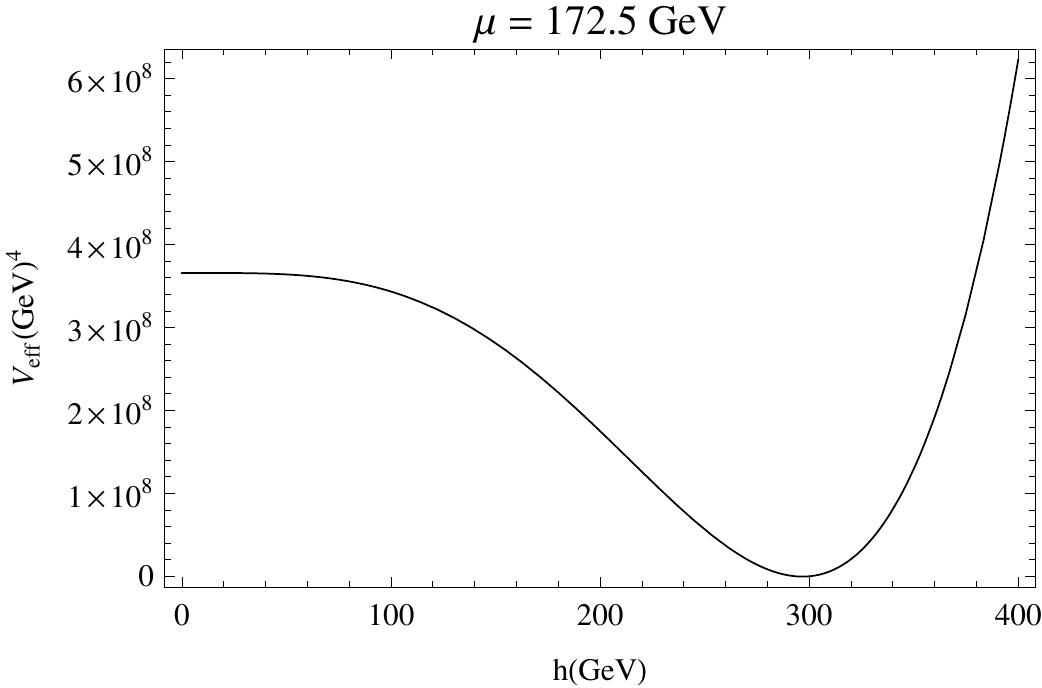}
\hskip20pt
\includegraphics[height=0.30\textwidth]{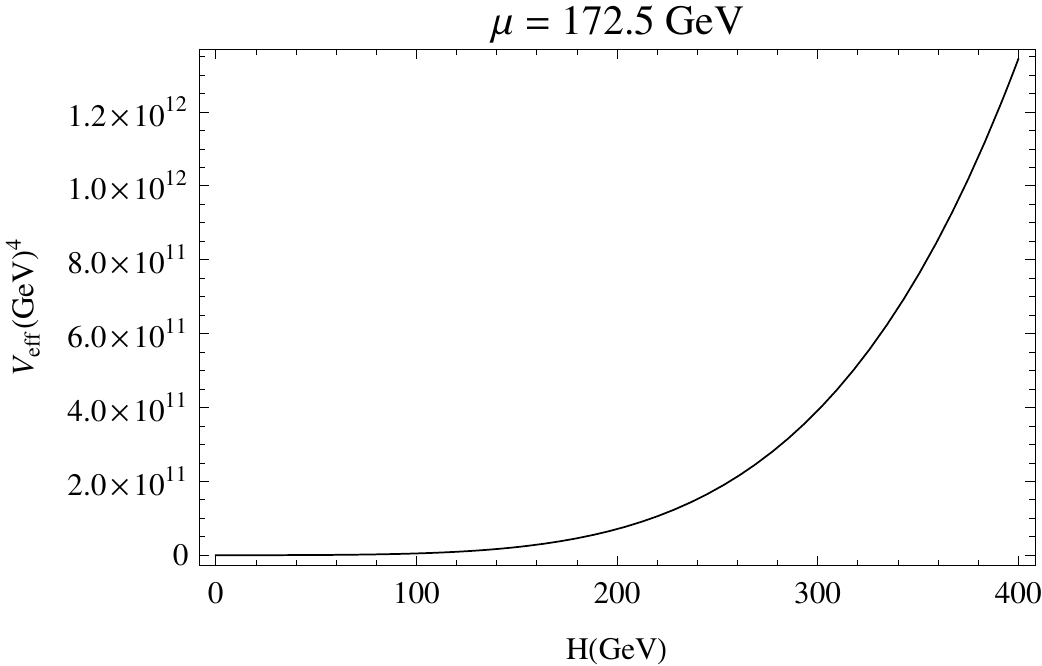}
\vspace*{0.2cm}
\hskip5pt  (a)
\hskip240pt  (b)
\caption{Effective potential the present universe}
\label{present_plot}
\end{center}
\end{figure}

\section{Generation of the scale invariant density perturbation}
We now turn to the question of the generation of density perturbations in the early era when $V_{eff}$
(\ref{eff_pot}) simplifies to the form,
\begin{eqnarray}
V_{inf}\sim V_c+\frac{\lambda_2}{4}\,H_0^4,
\label{Vinf}
\end{eqnarray}
where $V_c=3.66\times10^8\,\rm{GeV}^4$ and $\lambda_2= -0.5$.
The Hubble parameter in this era can be calculated from Eq~(\ref{Vinf}),
\begin{eqnarray}
H_{inf}&=&\frac{1}{\sqrt{3}}\frac{V_{inf}^{1/2}}{M_p}\sim\frac{1}{\sqrt{3}}\frac{V_0^{1/2}}{M_p}\\
&=&9.05\times10^{-16}\rm{GeV}
\end{eqnarray}

We take the inert Higgs doublet to be conformally coupled to gravity and  the action for this field can be written as,
\begin{eqnarray}
S=\int d^4x \sqrt{-g}\left[g^{\mu\nu}\partial_{\mu}H_2^{*}\partial_{\nu}H_2-\frac{R}{6}H_2^{*}H_2- V_{inf}\right]
\end{eqnarray}
where $H_2$ contains the neutral part of the inert doublet i.e. $H_2=\frac{H_0+i\, A_0}{\sqrt{2}}$ and
$R$ is the scalar curvature, which conformally
coupled with the field $H_2$. The equation of the field $H_2$ will be,
\begin{eqnarray}
\ddot{H_2}+\left(\frac{k}{a}\right)^2-3H\,\dot{H_2}+\frac{R}{6}H_2+\frac{\partial V_{inf}}{\partial H_2}=0
\label{eq1}
\end{eqnarray}
where $a$ and $H$ are the scale factor and Hubble constant respectively.
Now defining $H_2=\frac{\chi_{H_2}}{a}$ and rewriting the Eq~(\ref{eq1}), we will get,
\begin{eqnarray}
\chi_{H_2}^{''}+\left(k^2-\frac{a^{''}}{a}\right)\chi_{H_2}+\frac{R}{6}a^2\chi_{H_2}+a^3 \frac{\partial V_{inf}}{\partial H_2}=0
\label{eomH2}
\end{eqnarray}
where $\prime$ denotes the derivative with respect to conformal time $\eta$.
We note that both $\frac{a^{''}}{a}$ and $\frac{R}{6}a^2$ equal to $\frac{2}{\eta^2}$ and the two terms in Eq.~(\ref{eomH2}) cancel.
So the equation for $H_2$ becomes,
\begin{eqnarray}
\chi_{H_2}^{''}+k^2\chi_{H_2}+a^3 \frac{\partial V_{inf}}{\partial H_2}=0
\label{eom1}
\end{eqnarray}
Expressing $\chi_{H_2}=\rho\exp(i\,\theta)$, the conserved current will be of the form,
\begin{eqnarray}
\frac{d}{d\eta}(\rho^2\theta^{\prime})=0
\end{eqnarray}
Hence, the field rolls  along the radial direction while the phase $\theta$ remains constant with the increase of  $\rho$.
Without loss of generality we can choose the fixed phase such  that the field $H_2$ has only real component neutral
component $\chi_{H_0}$. The perturbations of $H_2$ will be along the imaginary axis and we can denote the full $H_2$
with the perturbations as $\chi_{H_2}=\chi_{H_0}+i\,\delta\chi_{A_0}$, from Eq~(\ref{eom1}) the equation of motion
of $\chi_{H_0}$  will be,
\begin{eqnarray}
\chi_{H_0}^{\prime\prime}+k^{2}\chi_{H_0}-\frac{\lambda_2}{2}\chi_{H_0}^3=0
\end{eqnarray}
Considering $k\ll 1/\eta$ at late time, the solution will be,
\begin{eqnarray}
 \chi_{H_0} \approx\frac{1}{\sqrt{-\lambda_2}(\eta_*-\eta)}
\label{chi_{H_0}}
\end{eqnarray}
where $\sqrt{-\lambda_2}$ is a real quantity as $\lambda_2$ is negative and $\eta_*$ is a constant of integration.
 At the end of inflation when $\mu<< 10^4$ the shape of the
potential changes, and $H_0$ starts rolling back to zero.

Starting from (16) we see that the equation of motion of the perturbation, $\delta\chi_{A_0}$ is given by,
\begin{eqnarray}
\delta\chi_{A_0}^{\prime\prime}+k^{2}\delta\chi_{A_0}+\frac{\lambda_2}{2}\chi_{H_0}^2\delta\chi_{A_0}=0
\end{eqnarray}
Substituting $\chi_{H_0}$ from Eq~(\ref{chi_{H_0}}), the equation becomes,
\begin{eqnarray}
\delta\chi_{A_0}^{\prime\prime}+k^{2}\delta\chi_{A_0}-\frac{1}{2(\eta_*-\eta)^2}\delta\chi_{A_0}=0
\end{eqnarray}
This equation can be solved for early times and later times separately.
At early time  $\left(k(\eta_* -\eta)\gg 1\right)$, third term can be neglected and the solution will be
\begin{eqnarray}
\delta\chi_{A_0}=\frac{1}{(2\pi)^{3/2}\sqrt{2k}}\exp^{ik(\eta_* -\eta)}
\end{eqnarray}
At later times, when $\left(k(\eta_* -\eta) \ll 1\right)$, third term will dominate and in this case solution will look like,
\begin{eqnarray}
\delta\chi_{A_0}\sim\frac{1}{k^{3/2}(\eta_* -\eta)}
\end{eqnarray}
Hence, the super-horizon perturbations of the phase can be defined as,
\begin{eqnarray}
\delta \theta \equiv \delta\chi_{A_0}/\chi_{H_0}
\end{eqnarray}
Therefore the perturbation of the phase $\delta \theta$ becomes,
\begin{eqnarray}
\delta \theta = \frac{\sqrt{-\lambda_2}}{k^{3/2}}
\end{eqnarray}
The power spectrum of $\delta \theta$ is scale invariant,
\begin{eqnarray}
{\cal P}_{\delta \theta}=\frac{k^3}{2\pi^2}|\delta\theta|^2=\frac{-\lambda_2}{2\pi^2}
\label{power}
\end{eqnarray}

If one considers the $k$ dependence of the equation of motion of $H_0$ as discussed in \cite{Libanov:2011hh} there will be a deviation
from the scale free power spectrum (\ref{power}) which will give rise to a non-zero spectral index,
\begin{eqnarray}
n_s -1 = \frac{3\lambda_2}{4\pi^2}
\end{eqnarray}
From Table-(\ref{table2}) we see that in the early universe $\lambda_2=-0.5$ which gives the spectral index $n_s-1= -0.04$,
which is consistent with the WMAP observation of $n_s=0.967 \pm 0.014$ \cite{WMAP}. The perturbations of the phase
$\delta \theta= \delta A_0/H_0$ can be converted to adiabatic perturbation by the decay of the $A_0$ field into standard
model fields as in the curvaton mechanism \cite{Curvaton}. The amplitude of adiabatic perturbation is related to the
phase perturbation as
\begin{equation}
P_\zeta =r^2 \frac{P_{\delta \theta}}{\theta_c^2}=r^2 \frac{-\lambda_2}{2 \pi^2 \theta_c^2}
\end{equation}
where $r$ is the ratio of the energy density in the $A_0$ field oscillations to the total energy density.
Taking the unperturbed phase to be $\theta_c\sim\pi/2$, and with $\lambda_2=-0.5$ we see that $r= 2 \times 10^{-4}$
to give the required $P_\zeta= 10^{-10}$.


\section{Scalar mass spectrum}
As the field $H_2$ has a zero vev in the present universe   the lightest neutral components of $H_2$  will be stable and
can be candidates for dark matter. We study the masses of the fields in present universe from the effective potential.
Taking $<H_1>\,=\,246 \,{\rm GeV}$ and $<H_2>\,=\,0 \,{\rm GeV}$ and for $\lambda_i$ as in Table-(\ref{table1}) we find the mass spectrum
of scalars in the present universe is as given in Table 3. We see that the field $A_0$ can be a candidate for heavy dark matter.
We also see that the Higgs mass is predicted to be $M_h=291$ GeV which is not ruled out \cite{Peskin} and may be observed at the LHC.

\begin{center}
\begin{table}[h]
\begin{tabular}{|c|c|c|c|c|}\hline
 $M_h$ & $M_{H^0}$  & $M_{A^0}$ & $M_{H^{\pm}}$  \\
\hline
$\,\,$291 $\,\,$& $\,\,$593$\,\,$& $\,\,$550 $\,\,$ & $\,\,$1228$\,\,$  \\
\hline
\end{tabular}
\caption{Scalar mass spectrum in GeV}
\end{table}
\end{center}

\section{Conclusions}
The inert Higgs doublet model gives is a natural extension of the standard model and can be used for explaining the
electroweak symmetry breaking by loop corrections \cite{Hambye:2007vf} starting from a scale invariant tree level potential.
We connect the scale invariance of the inert Higgs potential to the generation of scale invariant spectrum of a conformally
coupled  scalar as discussed by Rubakov and collaborators \cite{Rubakov:2009np,Libanov:2010nk,Libanov:2011hh}. The requirement
of scale invariance at high energy scale and electroweak symmetry breaking at low energies fixes the coupling constants of the
theory. Specifically we find that the the quartic coupling of the inert doublet, $\lambda_2=-0.5$ at $\mu=10^4$ GeV 
which predicts the spectral index of the power spectrum of the perturbations to be consistent with observations. The amplitude of
the power spectrum $P_\zeta$ can be tuned to be consistent with the observations by choosing a suitable curvaton mechanism. We
make predictions for masses of the Higgs bosons and the dark matter (which is the lightest neutral component of the inert doublet)
which can be tested in forthcoming experiments.


\end{document}